\documentclass[a4paper]{jpconf}
\usepackage{color}
\usepackage{graphicx}
\usepackage{float}

\newcommand{\singlet}{{}^1\mathrm{S}_0}
\newcommand{\triplet}{{}^3\mathrm{P}_2}
\newcommand{\Li}{{}^2\mathrm{S}_{1/2}}
\newcommand{\fLi}{${}^6$Li}
\newcommand{\bYb}{${}^{174}$Yb}
\newcommand{\ErYb}{E_{\mathrm R}^{\mathrm{Yb}}}
\newcommand{\ErLi}{E_{\mathrm R}^{\mathrm{Li}}}
\newcommand{\cm}{\mathrm{cm}}

\newcommand{\nm}{\mathrm{nm}}
\newcommand{\msec}{\mathrm{ms}}
\newcommand{\s}{\mathrm{s}}
\newcommand{\G}{\mathrm{G}}
\newcommand{\mG}{\mathrm{mG}}

\newcommand{\nK}{\mathrm{nK}}

\begin{document}

\title{Ultracold collisions in the Yb-Li mixture system}

\author{Florian Sch\"{a}fer$^1$, Hideki Konishi$^1$, Adrien Bouscal$^2$, Tomoya Yagami$^1$, Matthew D Frye$^3$, Jeremy M Hutson$^3$ and Yoshiro Takahashi$^1$}
\address{$^1$ Department of Physics, Graduate School of Science, Kyoto University, Kyoto 606-8502, Japan}
\address{$^2$ D\'{e}partement de Physique, \'{E}cole Normale Sup\'{e}rieure, PSL Research University, 24 rue Lhomond, 75005 Paris, France}
\address{$^3$ Joint Quantum Centre (JQC) Durham-Newcastle, Department of Chemistry, University of Durham, South Road, Durham, DH1 3LE, United Kingdom}
\ead{schaefer@scphys.kyoto-u.ac.jp}

\begin{abstract}
	We report our experimental results on the collisional physics between
	non-S-state atoms (ytterbium (Yb), effectively a two-electron system, in the
	metastable $\triplet$ state) and S-state atoms (lithium (Li), an alkali
	metal, in the ground state). At low magnetic fields, by measuring inelastic
	interspecies collisional losses in the double quantum degenerate mixture we
	reveal the strong dependence of the inelastic losses on the internal spin
	states of both species and suppressed losses in stretched state
	configurations. Increasing the magnetic field up to $800~\G$ we further
	investigate the magnetic field dependence of the collisional interactions.
	There, smoothly increasing inelastic losses are observed towards higher
	fields. The combined knowledge of both the magnetic field and the spin state
	dependence of the collisional losses of this prototypical mixture system of
	non-S-state and S-state atoms provides a significant step forward towards
	controllable impurity physics realized in the Yb-Li ultracold system.
\end{abstract}

\section{Introduction}
\label{sec:intro}

The understanding and control of the collisional properties of single-species
ultracold atomic systems was and is instrumental to their application in the
research of fundamental phenomena, quantum simulation and quantum computation.
For example, elastic collisions enable us to reach ultracold temperatures via
forced evaporative cooling~\cite{ketterle_evaporative_1996}, light-assisted
collisions lead in photo-association processes to molecular
states~\cite{weiner_experiments_1999}, manipulation of the scattering lengths
by means of Feshbach resonances~\cite{chin_feshbach_2010} facilitate the
crossover from Bose-Einstein condensates to Bardeen–Cooper–Schrieffer
superfluids~\cite{bourdel_experimental_2004, zwierlein_condensation_2004} and
give insight to Efimov trimer physics~\cite{zaccanti_observation_2009}, and
controlled collisions between atoms form one possible basis for quantum gates
in quantum computation schemes~\cite{daley_quantum_2011}.

Going from single-species to two-species systems adds additional degrees of
freedom to the experimental toolkit. For example, the physics of impurity
systems leads to questions of both transport and localization of quantum
matter~\cite{massignan_polarons_2014, chien_quantum_2015,
nishida_transport_2016}. Related questions in mixed dimensional systems range
from basic phase diagram considerations~\cite{nishida_universal_2008} to the
possibility of chiral p-wave superfluids~\cite{nishida_induced_2009,
wu_topological_2012} and their application to topologically protected quantum
computation~\cite{kitaev_fault-tolerant_2003, tewari_quantum_2007}. Also,
polar molecules associated from two-species ultracold mixtures allow
implementation of novel spin-lattice models as new platforms for quantum
computation protocols~\cite{micheli_toolbox_2006}.

\begin{figure}[tb]
	\centering
	\includegraphics[width=8cm]{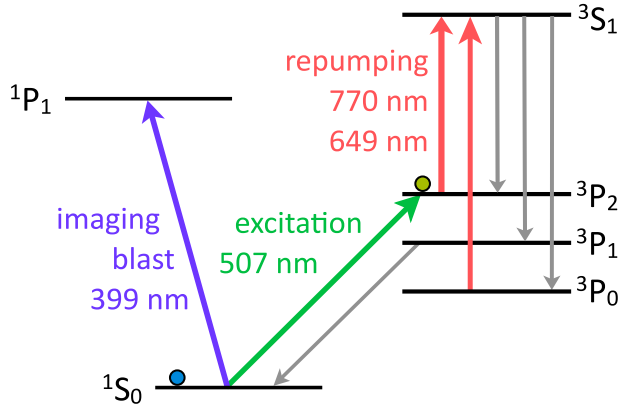}\hspace{2pc}%
	\begin{minipage}[b]{7cm}
	\caption{Yb level scheme of relevance to the experiments. Light at $507~\nm$
		connects the ground to the metastable $\triplet$ state (green arrow).
		Fluorescence imaging at the strong $\singlet - {}^1\mathrm{P}_1$
		transition (purple arrow) reveals the remaining $\triplet$ atoms after a
		blast of any remaining ground state atoms and transfer of the excited
		atoms to the ground state by two repumping lasers (red arrows) followed by
		spontaneous decay (gray arrows).}
	\label{fig:levelscheme}
\end{minipage}
\end{figure}

We here present and summarize our experiments on quantum degenerate mixtures
of fermionic \fLi\ and bosonic \bYb, where Yb is in the metastable $\triplet$
state (see Fig.~\ref{fig:levelscheme}). First, at low magnetic bias fields we
investigate inelastic interspecies collisional losses in the double quantum
degenerate mixture that reveal the dependence of the inelastic losses on the
internal spin states of both species. As a result, we discover two very
distinct regimes: in collisions with \fLi\ in the $F = 1/2$ hyperfine state
overall constant inelastic collision rates are found. These results are in
agreement with our previous report~\cite{konishi_collisional_2016}. The
collisional system with \fLi\ in the $F = 3/2$ state, however, exhibits quite
a different behavior. Now the magnetic sublevel of \bYb\ matters and we
observe suppressed losses for stretched state configurations and up to
40-times increased relaxation rates in the other cases. Second, we measure the
magnetic field dependence of the inelastic collision rates between \bYb, again
in the metastable $\triplet$ state, and \fLi\ for several hyperfine levels of
Li. Those experiments reveal a smooth but steady increase of the inelastic
collision rates up to the highest investigated magnetic fields of $800~\G$ and
underline the importance of choosing collisional channels with suppressed
inelastic dynamics.

This Progress Report is organized as follows: After this introduction we will
review the necessary experimental details in Sec.~\ref{sec:experiment}. The
individual results are then presented and explained in
Sec.~\ref{sec:spin_dependence} and Sec.~\ref{sec:field_dependence}.
Section~\ref{sec:conclusions} concludes the Report with a summary and a
general discussion of our experimental findings.

\section{Experiment}
\label{sec:experiment}

We here review the necessary experimental methods that are also discussed in
more depth and detail in the related works~\cite{hara_quantum_2011,
schafer_spin_2017, schafer_spectroscopic_2017}. Starting from a hot beam of
thermal Yb and Li atoms from a dual-species oven, both species are
sequentially Zeeman slowed and trapped in a magneto-optical trap (MOT). The
atomic beam contains abundances of all stable isotopes of both species. By
proper selection of the laser frequencies we choose to to slow down and cool
only selected isotopes, namely bosonic \bYb\ and fermionic \fLi. From the MOT
the pre-cooled samples are transferred to a crossed far-off-resonant trap
(FORT) created by two focused laser beams at wavelengths of about $1060~\nm$.
Next, the intensities of the FORT lasers are decreased as to drive forced
evaporative cooling of the \bYb\ atoms until quantum degeneracy, a
Bose-Einstein condensate (BEC), is reached. Our setup is not designed to
perform evaporative cooling of Li independently. Instead, Li is in thermal
contact with Yb and is cooled to quantum degeneracy via sympathetic
cooling~\cite{myatt_production_1997}.

During the initial phase of the evaporative cooling ramp we apply a
$0.5~\msec$ pulse of circularly polarized light resonant to either the Li $F =
1/2 \rightarrow F' = 1/2$ or the $F = 3/2 \rightarrow F' = 3/2$ D1-line
transition. Together with a repumping light derived from the D2-line MOT light
lasers we can effectively prepare a spin-polarized \fLi\ sample in either of
the four $F=1/2$, $m_F=\pm1/2$ or $F=3/2$, $m_F=\pm3/2$ ground states. This
polarization state remains with high fidelity until quantum degeneracy is
reached and time-of-flight absorption imaging after application of a strong
magnetic field gradient reveals above 90\% of final spin-polarization. By
choosing appropriate loading durations for the \bYb\ and \fLi\ MOTs at this
point we typically obtain a BEC of $10 \times 10^4$ \bYb\ atoms and a Fermi
degenerate \fLi\ gas of $3 \times 10^4$ atoms. The temperature of Li is about
$300~\nK$ or about $0.2\ T_{\rm F}$, where $T_{\rm F}$ is the Fermi
temperature. Typical densities at this point are $5 \times
10^{14}~\cm^{-3}$ for \bYb\ and $5 \times 10^{12}~\cm^{-3}$ for \fLi.

\begin{figure}[tb]
	\centering
	\includegraphics[width=8cm]{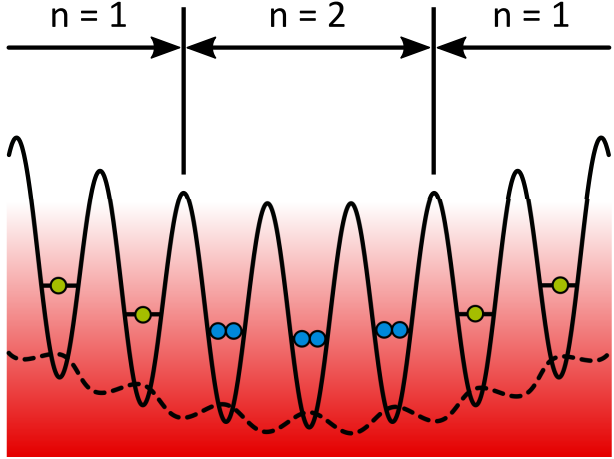}\hspace{2pc}%
	\begin{minipage}[b]{7cm}
	\caption{Quantum degenerate mixture of \bYb\ and \fLi\ in a 3D optical
		lattice. Yb experiences a deep lattice potential (solid line) and forms a
		Mott-insulator shell structure (blue and green dots, only occupations
		numbers $n \le 2$ shown). The effective potential for Li is shallow
		(dashed line) and the atoms remain unlocalized (red shaded area). For our
		investigation of inter-species scattering we focus on the interaction
		between \fLi\ and \bYb($\triplet$) atoms (green dots) in $n = 1$ sites.}
	\label{fig:lattice}
\end{minipage}
\end{figure}

As we want to study the collisional dynamics between non-S-state atoms (\bYb\
in the metastable $\triplet$ state) and S-state atoms (spin-polarized \fLi\ in
the ground state) it is necessary (i) to prepare the Yb atoms in the
$\triplet$ state and (ii) to prevent fast losses stemming from inelastic
Yb($\triplet$)-Yb($\triplet$) collisions~\cite{uetake_spin-dependent_2012}. To
both ends we adiabatically load the quantum degenerate mixture into a
three-dimensional (3D) optical lattice, formed by three pairs of
counter-propagating laser beams of wavelength $\lambda_{\rm L} = 532~\nm$. At
$\lambda_{\mathrm L}$ the sign of the polarizabilities of Li and Yb are
opposite. The chosen lattice depths are about $15\, \ErYb$ for \bYb\ and
$0.7\, \ErLi$ for \fLi, where $\ErYb$ ($\ErLi$) is the recoil energy of Yb
(Li). Hence, while \fLi\ only experiences a minor modulation of its wave
function \bYb\ forms a Mott-insulator state (see Fig.~\ref{fig:lattice} and
Refs.~\cite{bloch_many-body_2008, konishi_collisional_2016}). The difference
in gravitational sag of the two species is addressed by an additional light
field gradient, increasing the otherwise small spatial overlap between \bYb\
and \fLi~\cite{konishi_collisional_2016}.

Using pairs of magnetic coils in Helmholtz configuration we then apply the
desired bias fields, typically between $100~\mG$ and $800~\G$. There, the
metastable $\triplet$ state of Yb experiences significant Zeeman splitting of
$2.1~{\rm MHz}/\G \times m_J$. By application of a $0.5~\msec$ long pulse of
507-nm light resonant to the $\singlet$ - $\triplet$ transition we then site
and state selectively excite \bYb\ atoms only in singly occupied lattice sites
($n = 1$) and for a desired target Zeeman state $-2 \le m_J \le 2$. This
effectively suppresses Yb($\triplet$)-Yb($\triplet$) collisional events. The
excitation strength is chosen such that the number of excited \bYb\ atoms is
about 10\% of the total number of \fLi\ atoms. In the following we can
therefore assume that there is only a small number of non-S-state \bYb\ atoms
immersed within a large Fermi sea of S-state \fLi\ atoms. After the excitation
a short light pulse at 399 nm removes the remaining ground state \bYb\ atoms
from the system, ensuring no contributions from ground state Yb atoms to our
results. During a variable holding time we then allow the excited \bYb\ atoms
to interact with the background \fLi\ gas. We note that losses of
excited-state atoms due to collisions with hot background gas atoms and
scattering from the trap lasers limit the lifetime of the $\triplet$ atoms to
$(850 \pm 300)~\msec$ which is much longer than the typical timescales
involved in interactions with \fLi.

In the detection stage of the experiment we measure the remaining
\bYb($\triplet$) atoms via the very sensitive fluorescence imaging method on
the strong 399-nm transition (see Fig.~\ref{fig:levelscheme} and
Ref.~\cite{konishi_collisional_2016}). For this we first apply a 399-nm light
pulse to remove any accumulated \bYb($\singlet$) atoms from the system,
followed by a short pulse of two repumping lasers (cf.\
Fig.~\ref{fig:levelscheme}) that transfers the remaining \bYb($\triplet$)
atoms back to the ground state. There they are recaptured by a MOT operating
on the 399-nm transition whose fluorescence light is integrated for $1~\s$ on
a sensitive charge-coupled device (CCD) camera to obtain a high-contrast
signal of the MOT proportional to the number of \bYb($\triplet$) atoms.
Imaging of the \fLi\ atoms proceeds via standard time-of-flight absorption
imaging.

\section{Spin dependence}
\label{sec:spin_dependence}

A first series of experiments is to establish the dependence of the inelastic
collisions on the spin state of both the non-S-state \bYb($\triplet$) and the
S-state \fLi($\Li$) collisional partners. This is to gain more insight into
anisotropy induced relaxation processes~\cite{reid_fine-structure_1969,
mies_molecular_1973, krems_suppression_2005} in this collisional process.

\begin{figure}[tb]
	\centering
	\includegraphics{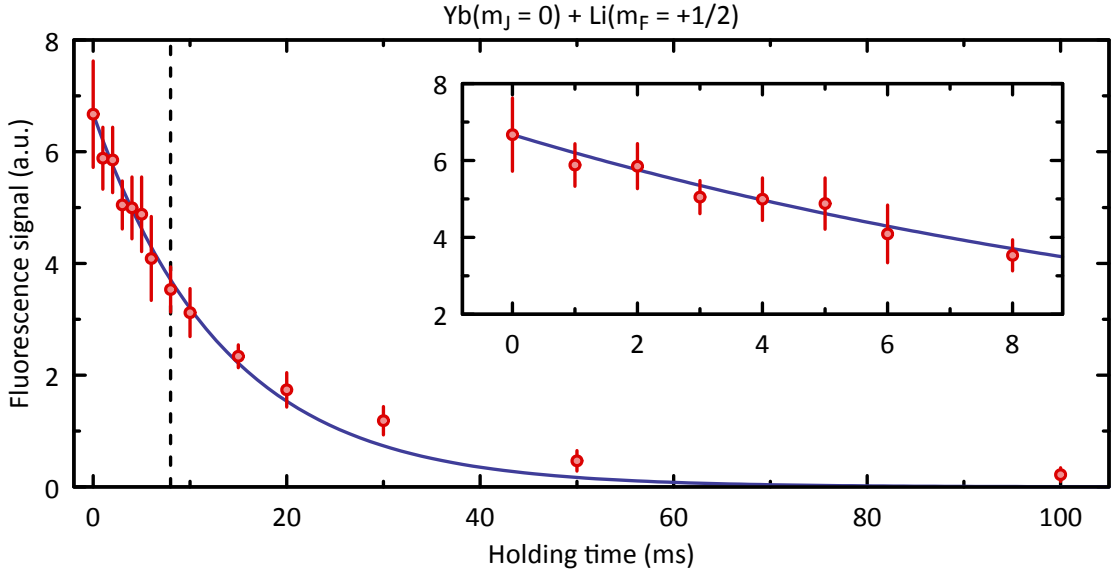}
	\caption{Decay of $^{174}{\rm Yb}(\triplet, m_J = 0)$ excited state atoms
		through inelastic collisions with $^{6}{\rm Li}(\Li, F = 1/2, m_F = +1/2)$
		ground state atoms, where we show the intensity of the obtained
		fluorescence signal as a function of the holding time. The data points
		(red) are the mean of independent measurements at each holding time. The
		error bars represent their standard deviations. An exponential decay
		(blue, solid line) is fitted to the initial time window (up to the dashed
		line) and shown enlarged in the inset. The lifetime obtained here
		is $(13.5 \pm 1.2)~\msec$ corresponding to an inelastic collisional rate
		coefficient of $(4.6 \pm 2.2) \times 10^{-11}~\cm^3/\s$.}
	\label{fig:decay}
\end{figure}

The experiment proceeds as outlined in Sec.~\ref{sec:experiment}. During the
first $100~\msec$ of the adiabatic increase of the optical lattice depth we
set a homogeneous bias magnetic field of $200~\mG$. This is to define a good
quantization axis and to lift the degeneracy of the five $m_J = -2, \ldots,
+2$ Zeeman states of \bYb($\triplet$). The experiment is executed for all
combinations of Yb Zeeman states and accessible Li ground states, 20
configurations in total. For each combination we repeat the experiment several
times, varying each time the holding time of the mixture in the optical
lattice up to $100~\msec$. We point out that at the present bias magnetic
field we could confirm that our method is sensitive to fine-structure changing
collisions (e.g.\ ${\rm Yb}(^3{\rm P}_2) \rightarrow {\rm Yb}(^3{\rm P}_1)$)
and hyperfine-structure changing collisions ($F = 3/2 \rightarrow F = 1/2$).
In contrast, spin-changing collisions ($m_J$ or $m_F$ change) do not lead to
sufficient gain in kinetic energy for the atoms to leave the optical
lattice~\cite{schafer_spin_2017}.

A representative decay of the remaining $\triplet$ atoms is shown in
Fig.~\ref{fig:decay}. The data qualitatively can be divided into two decay
regimes. An initial, faster decay is followed by a slower decay dynamics.
Having confirmed the linearity of our fluorescence light detection system, we
interpret this as the crossover from a decay mostly governed by interspecies
collisions to a mixed dynamics where also collisions of the Yb($\triplet$)
atoms with background gas atoms combined with slightly reduced Li densities
gain importance. We therefore typically limit the analysis of the data to the
first few milliseconds (indicated by the dashed line and the inset in
Fig.~\ref{fig:decay}). The initial decay is then analyzed in terms of rate
equations taking Yb($\triplet$)-Li inelastic losses and Yb($\triplet$)
one-body losses into account. The details of the numerical analysis are
described in our related work~\cite{schafer_spin_2017}. The analysis then
gives access to a lifetime $\tau$ that, with knowledge of the atomic densities
and their overlap, is translated into the inelastic loss parameter.

\begin{figure}[tb]
	\centering
	\includegraphics{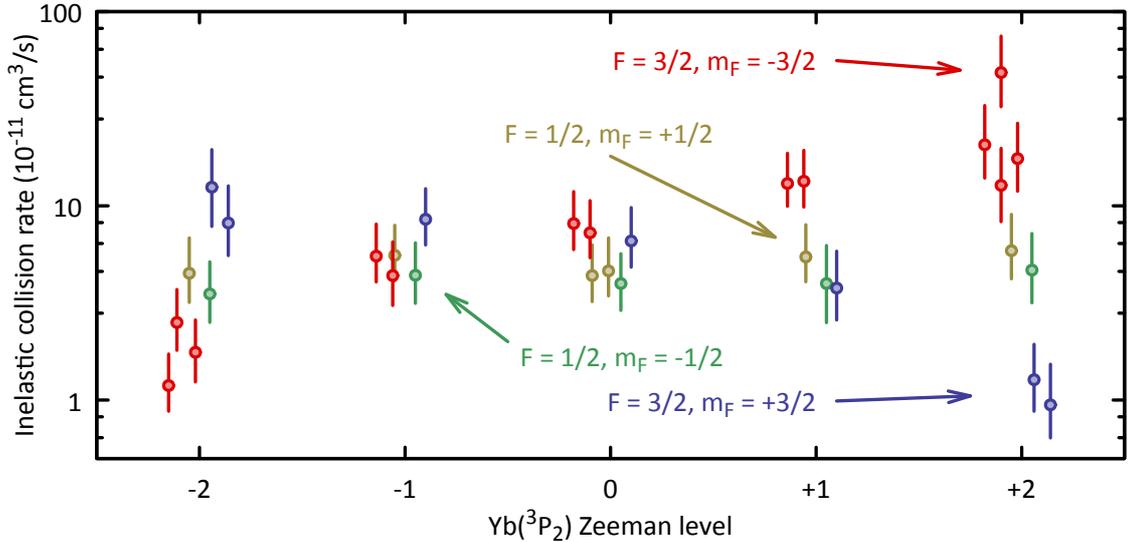}
	\caption{Inelastic collision rates between metastable
	\bYb($\triplet$) and ground state \fLi\ in various hyperfine states at
	$200~\mG$ bias magnetic field. Inelastic collisions with $^{6}{\rm Li}(\Li,
	F = 1/2, m_F=\pm 1/2)$ (yellow and green symbols) show only a weak
	dependence on the $m_J$ magnetic state of \bYb. In contrast, a clear
	$m_J$-state dependence in collisions with $^{6}{\rm Li}(\Li, F = 3/2,
	m_F=\pm 3/2)$ is observed (blue and red symbols). The data points represent
	the median values, the error bars the $1-\sigma$ confidence interval of the
	determined collision rates. The data are slightly shifted
	horizontally for better visibility.}
	\label{fig:rates}
\end{figure}

Figure~\ref{fig:rates} summarizes the results of this survey. The data fall
into two categories. Looking at intermediate collisional rates we observe that
the data for collisions with \fLi($\Li, F = 1/2, m_F = \pm 1/2$) all have very
similar inelastic loss rates of about $4 \times 10^{-11}~\cm^3/\s$. Collisions
with \fLi\ in the $F = 3/2$ hyperfine state, however, show a very pronounced
and systematic dependence on the magnetic sublevel of both \bYb($\triplet$)
and \fLi. Especially for collisions between \bYb($\triplet, m_J = +2$) and
\fLi($\Li, F = 3/2, m_F = \pm 3/2$) a 40-fold change of the loss rate is
observed.

The findings are discussed in terms of the interspecies interaction operator
$\hat{U}$ described in detail for the Yb($\triplet$)-Li($\Li$) system
in~\cite{gonzalez-martinez_magnetically_2013}. The anisotropy introduced into
$\hat{U}$ by the non-S-state species leads to significant differences in the
potentials of the four possible Yb($\triplet$)-Li($\Li$) molecular
states~\cite{petrov_magnetic_2015} and drives strong, spin dependent
relaxation processes~\cite{krems_suppression_2005}. Indeed our inelastic loss
rates are generally several orders of magnitude larger than those reported for
systems where the anisotropy is suppressed by screening effects of outer
s-shell orbitals~\cite{hancox_suppression_2005}. The efficient suppression of
inelastic losses observed in the configurations \bYb($\triplet, m_J =
+2$)-\fLi($\Li, F = 3/2, m_F = +3/2$) and \bYb($\triplet, m_J =
-2$)-\fLi($\Li, F = 3/2, m_F = -3/2$) stems from the lack of allowed inelastic
decay channels for these stretched state configurations, assuming conservation
of the total electronic angular momentum and its projection. According
to~\cite{krems_electronic_2004} one can identify two contributions to
$\hat{U}$, internal and external anisotropy. The internal anisotropy conserves
the total electronic angular momentum and its projection. It therefore cannot
cause inelastic changes in stretched state situations. In contrast, the
external anisotropy couples to the rotational momentum of the nuclei and does
not abide by the above conservation rule. In this light, our stretched state
results allow direct access to the contribution by the external anisotropy
that with inelastic collision rates of about $10^{-11}~\cm^3/\s$ is in good
agreement with earlier predictions~\cite{gonzalez-martinez_magnetically_2013}.
The impact of the internal anisotropy is expressed by our findings for the
non-stretched collisional systems where losses increase by one order of
magnitude.

In order to understand the patterns of inelastic collision rates, we perform
coupled-channel scattering calculations on Li+Yb($\triplet$). We use the same
methods as described in Ref.~\cite{gonzalez-martinez_magnetically_2013},
with all three spin-orbit states of Yb(${}^3\mathrm{P}$) included. The
interaction potential for this system is known only from electronic structure
theory~\cite{gopakumar_relativistic_2010} and is not accurate enough for
quantitative predictions of scattering lengths or resonance predictions, but
it should adequately represent the qualitative behaviour of the system. In
order to sample the effects of different scattering lengths for the different
electronic states, we construct 10 different interaction potentials. Each of
these potentials is composed of 4 potential curves ($^2\Sigma^+$, $^2\Pi$,
$^4\Pi$, and $^4\Sigma^+$), each of which is multiplied by a random scaling
factor that varies over a range sufficient to randomise the scattering length
of that potential completely.

We calculate loss rates at $200~\mG$ for all 20 combinations of states for the
experimental results shown in Fig.~\ref{fig:rates}, for all 10 of our random
potentials. There is a large variation of about 2 orders of magnitude in the
overall magnitude of the loss rates between different potentials. This range
encompasses the experimental results. For most interaction potentials, there
is an approximate symmetry between calculated loss rates for $(m_F, m_J)$ and
$(-m_F, -m_J)$, although there are often differences of up 20\%. Among the
states of Li with $F=1/2$, the differences between the loss rates for
different Zeeman components are generally small, in agreement with the
experiment.

For several of the potentials there is strong suppression of the loss rates
for the spin-stretched states. Losses from these states require a change in
$L$, the quantum number for end-over-end rotation, and so depend on the
external anisotropy. In one view, this anisotropy is proportional to the
difference between $\Sigma$ and $\Pi$ potentials. However, there may sometimes
be a coincidental cancellation of the couplings, leading to suppression of
loss from these states for some of our potentials. This cancellation may be
conceptually similar to alkali-alkali systems with similar singlet and triplet
scattering lengths~\cite{julienne_collisional_1997, burke_impact_1997}, where
spin exchange is suppressed, but the details in this case are likely to be
much more complicated. From the experimental results, it appears that the real
system also exhibits this suppression. 

\section{Magnetic field dependence}
\label{sec:field_dependence}

We now turn our attention to the magnetic field dependence of the inelastic
loss rates. Field strengths up to $800~\G$ are considered. For field strengths
$> 300~\G$ we directly ramp up the field, similar to the previous experiment,
to the final value while preparing the optical lattice. For magnetic fields
below $300~\G$, due to technical constraints of our setup, direct excitation
is not efficiently possible. Instead we prepare and excite the sample at
$300~\G$ and then ramp down the field within $0.3~\msec$ to the desired target
field. Due to the strong Zeeman splitting of the Yb($\triplet$) state, the
inevitable technical noise of the magnetic field and the narrow linewidth of
the transition, we can reliably excite only the \bYb($\triplet, m_J=0$) in
this situation. Therefore, we limit this experiment to the $m_J=0$ Zeeman
state of \bYb($\triplet$). The data analysis is as in the low-field
measurements of the preceding section.

\begin{figure}[tb]
	\centering
	\includegraphics{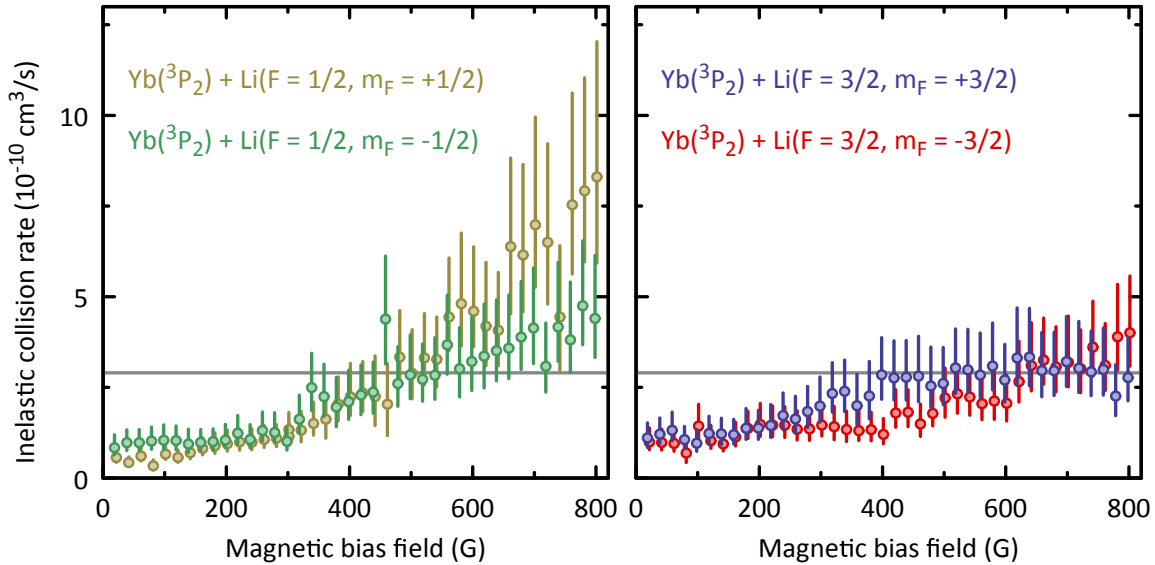}
	\caption{Summary of inelastic collision rates between $^{174}{\rm
		Yb}(\triplet, m_J = 0)$ and $^{6}{\rm Li}(\Li)$ state atoms for bias
		magnetic fields between $20$ and $800~\G$. The \fLi\ atoms are either
		prepared in one of the two available $F = 1/2$ states (left panel) or
		the accessible $F = 3/2$ states (right panel). For easier comparison
		both panels are drawn to the same scale. Also indicated is the expected
		universal rate (gray lines). In all cases a smooth increase of the
		inelastic collision rate is found. No strong, resonance-like features
		indicative of possible Feshbach resonances are observed. The data is
		slightly offset horizontally for better readability.}
	\label{fig:highfield}
\end{figure}

The results are summarized in Fig.~\ref{fig:highfield}. Note, that even in
this high-field situation we identify the states of Li by their low-field
quantum numbers ($F$, $m_F$) at which they were prepared and that are
adiabatically connected to the projections of the electronic and nuclear spin
angular momenta ($m_S$, $m_I$) at high magnetic field. By using non-stretched
state configurations the data include effects from both internal and external
anisotropy and therefore should give a good overview of the general high-field
behavior of the inelastic collisional loss rates. For all investigated spin
states of \fLi\ we find a smooth and general increase of the losses, starting
with low fields at about $10^{-10}~\cm^3/\s$ and staying below
$10^{-9}~\cm^3/\s$ when reaching $800~\G$. Especially for fields beyond about
$600~\G$ collisions involving \fLi\ in the $F = 3/2$ hyperfine state have
slightly lower losses then those with \fLi\ in the $F = 1/2$ state. Overall,
we find larger inelastic collisional rates than what was predicted
in~\cite{gonzalez-martinez_magnetically_2013} for collisions involving
Yb($\triplet, m_J = -2$). Also, for higher magnetic fields the losses approach
and go beyond the predicted universal
rate~\cite{gonzalez-martinez_magnetically_2013} of $2.9 \times
10^{-10}~\cm^3/\s$ (gray lines in Fig.~\ref{fig:highfield}); this corresponds
to a model in which incoming flux can be reflected from the long-range
potential, but all the flux that reaches short range is irreversibly lost with
unit probability.~\cite{idziaszek_universal_2010}. This again exemplifies the
importance of the anisotropy in $\hat{U}$ for ultracold collisions of
\bYb($\triplet$) and \fLi($\Li$).

We have carried out further coupled-channel calculations to understand these
results. We use the same interaction potentials as in the previous section,
but now perform calculations up to $1000~\G$ for the states shown in
Fig.~\ref{fig:highfield}. The general behavior of the loss rates for these
states is to increase from a small value at low field towards the universal
loss rate~\cite{idziaszek_universal_2010}, sometime rising above it at higher
fields. There is resonant structure in most cases, which creates large peaks
in loss rate. Some of which can have widths of hundreds of Gauss and are
responsible for rates higher than universal. Many of the resonant features
appear consecutively in the three lowest spin channels for Li ($F=1/2,
m_F=1/2$; $F=1/2, m_F=-1/2$; $F=3/2, m_F=-3/2$; all of which correspond to
electron spin $m_s=-1/2$) at progressively higher field, most likely due to a
single state cutting upwards in energy through each of the thresholds in turn.
Turning to the experimental results, the loss rates in the lowest Li channel
($F=1/2\, m_F=1/2$) rise strongly towards the upper end of the field range.
The loss rates in the next states up rise less strongly, but one of them is
still clearly above the universal rate. This behavior may be evidence of a
broad resonance in the loss rate, centered above $800~\G$ for the $F=1/2,
m_F=1/2$ channel, and at even higher fields for the higher $F=1/2, m_F=-1/2$
and $F=3/2, m_F=-3/2$ channels. Further measurements at higher fields would
be able to confirm if this is indeed the case.

\section{Conclusions}
\label{sec:conclusions}

We investigated the inelastic rates in the \bYb($\triplet$)-\fLi($\Li$) system
of non-S-state and S-state collisional partners. In contrast to S-state-only
collisions effects of anisotropy become of significant importance here. This
is evidenced by the strong dependence on the spin states we observed at low
magnetic fields. At higher magnetic fields we find inelastic collision rates
beyond the universal rate of complete loss at short distances. Furthermore
only a smooth variation of the loss rates is found. While the general findings
are supported by our coupled-channel calculations, the details are as yet
mostly beyond our understanding. This renders experimental investigations all
the more important and should encourage further theoretical studies. It would
for example be interesting to extend our high-field studies to the numerically
more investigated case of \bYb($\triplet, m_J = -2$) that might prove to be
more collisionally stable. 

In future experiments towards controllable impurity physics in the ultracold
Yb-Li mixture system consideration of the anisotropy and its associated
additional losses is necessary and stretched-state configurations should be
preferred where possible to suppress unnecessary loss mechanisms. Our work
thus provides a significant step towards a better understanding and control of
ultracold Yb($\triplet$)-Li mixtures.

\ack 

We thank Ruth Le Sueur for useful early discussions.
This work was supported by the Grant-in-Aid for Scientific Research of JSPS
Grants No.\ JP25220711, No.\ JP26247064, No.\ JP16H00990, No.\ JP16H01053,
No.\ JP17H06138, No.\ 18H05405, No.\ 18H05228, JST CREST Grant No.\
JPMJCR1673, the Impulsing Paradigm Change through Disruptive Technologies
(ImPACT) program by the Cabinet Office, Government of Japan, and MEXT-QLEAP.
Further support was provided by U.K.\ Engineering and Physical Sciences
Research Council (EPSRC) Grant No.\ EP/P01058X/1. HK achnowledges support from
JSPS.

\section*{References}

\providecommand{\newblock}{}

\end{document}